\documentclass{amsart}
\usepackage{graphicx}
\usepackage{amsmath,amsthm}
\usepackage{amssymb}
\usepackage{caption}
\usepackage[latin1]{inputenc}
\usepackage{enumerate}

\theoremstyle{plain}
\newtheorem*{lemacero}{Lema 0}
\newtheorem*{thm*}{Teorema}
\newtheorem*{cor*}{Corolario}
\newtheorem*{lema*}{Lema}
\newtheorem*{prop*}{Proposición}
\theoremstyle{definition}
\newtheorem*{defi*}{Definición}
\newtheorem*{Ej*}{Ejemplo}
\newtheorem*{obs*}{Nota}
\newtheorem*{Obs*}{Notas}

\newcommand{\cC}{{\mathcal C}}
\numberwithin{equation}{section}

\renewenvironment{abstract}
 {\small
  \begin{center}
  \bfseries \abstractname\vspace{-.5em}\vspace{0pt}
  \end{center}
  \vskip .3cm
  \list{}{
    \setlength{\leftmargin}{.5cm}%
    \setlength{\rightmargin}{\leftmargin}%
  }%
  \item\relax}
 {\endlist}

\begin{document}

\title[On the foundations of mechanics]
 {On the foundations of mechanics}

\author{R. J.  Alonso-Blanco y J.  Mu\~{n}oz-D{\'\i}az}

\address{Departamento de Matem\'{a}ticas, Universidad de Salamanca, Plaza de la Merced 1-4, E-37008 Salamanca,  Spain.}
\email{ricardo@usal.es, clint@usal.es}

\maketitle

\begin{abstract}
This note is an extended version of ``A note on the foundations of Mechanics'', arXiv: 1404.1321 [math-ph]. A presentation of its contents was given in a talk  in memorial homage to the professor Juan B. Sancho Guimerá. For this reason, it was written in spanish language. The matter of the note is a systematic foundation of the most classical part of Mechanics. The content by sections is:
\vskip .2cm
\begin{enumerate}
\item[0)] {\bf Notions and basic results}: in this section it is formulated the  fundamental lemma of Mechanics, Lemma 0, which  is the statement of the Newton's equation in arbitrary configuration space, and the field equation for its intermediate integrals. Some consequences are derived: determination of the force law from a family of trajectories, geometric formulation of constrained systems, and some remarks on reversibility or not of classical mechanical systems.
\item[1)] {\bf Conservative systems}: from Lemma 0 are directly derived the Canonical Hamilton equations, Maupertuis' principle, Hamilton-Jacobi equation, and the Schrö\-din\-ger equation for the lagrangian and conservative intermediate integrals; finally, we give the presentation of a conservative system as a projection of a geodesic system in higher dimension.
\item[2)] {\bf Time. Time constraints}: in a general mechanical system can not exist a ``time'' parameterizing all of the  trajectories; time must be imposed as a constraint; this constraint modifies the previous equations in a well defined way; as an example, all conservative system can be obtained as the result of a time constraint on a geodesic system. Hamilton-Jacobi equation for a system with a time constraint can be derived from a modification of the symplectic structure of the phase space; from the harmonic solutions of the Hamilton-Jacobi equation, are deduced wave  functions which hold Schrödinger equation.
\item[3)] {\bf Proper time. Relativistic forces}:  the only natural way of defining what a relativistic system must be in general mechanics is the property of its trajectories of being parameterizable by the proper time (which is the length for the given metric). Such relativistic systems are characterized by the fact of having a force form which belongs to the contact system, a condition which does not depend on the metric. The above said clarifies the role of antisymmetric  tensors in Relativity. All mechanical system admits a canonical relativistic correction.
\item[4)] {\bf Electromagnetic fields}: as it is known, an electromagnetic field is a closed 2-form in the given configuration space; this tensor field and the given metric  assign canonically a Lorentz force (like in every relativistic system); the problem is if the current defined by the first couple of Maxwell equations is an intermediate integral of the Lorentz force; this condition, in addition of being lagrangian for the symplectic structure modified by the electromagnetic field, gives the Klein-Gordon equation.
\item[5)] {\bf On the Hamilton-Noether Principle}: in our presentation of Mechanics, the point of departure is the general Newton equation as stated in Lemma~0 and the Variational Principles are deduced from it; this approach has the advantage of no need for justify these principles. For dissipative systems, the Hamilton Principle is deduced from a modification of the symplectic structure produced by a time constraint on a conservative system.
\item[6)] {\bf Schrödinger equation}: the Schrödinger equation for an intermediate integral of a conservative system was obtained from the very restrictive hypotheses of being lagrangian and conservative; in this section we propose the conditions that seem to be reasonably the most general ones within the Classical Mechanics for what may be interpreted as matter waves; we arrive to an equation which contains a non linear term added to the quantum Schrödinger equation; this term vanishes when the squared root of the density is a harmonic function.
\end{enumerate}
\end{abstract}
\bigskip

\setcounter{section}{-1}

\section{Nociones y resultados básicos}

Sean $M$ una variedad diferenciable, $TM$ su fibrado tangente.
Diremos que un vector o un campo tangente a $TM$ es \emph{
vertical} cuando aniquile al subanillo $\cC^\infty(M)$ de
$\cC^\infty(TM)$. Una 1-forma en un punto o una 1-forma en $TM$ se
dirá que es  \emph{ horizontal} cuando sea incidente con todos los
vectores verticales. En coordenadas locales $(x^1,\dots,x^n,\dot
x^1,\dots,\dot x^n)$ en $TM$, los vectores o campos verticales son
combinaciones lineales de los $\partial/\partial\dot x^i$, y las
1-formas horizontales, de las $dx^i$.

Cada 1-forma horizontal $\alpha$ define una función $\dot\alpha$
en $TM$, por la regla $\dot\alpha(v_a)=\langle \alpha,v_a\rangle$,
para cada $v_a\in TM$. En particular, para cada
$f\in\cC^\infty(M)$, la función $\dot{\overline{df}}$ se denotará
$\dot f$. La aplicación
$$\dot d\colon\cC^\infty(M)\to\cC^\infty(TM),\quad f\mapsto \dot
d(f)=\dot f$$ es, esencialmente, la diferencial. Puede pensarse
$\dot d$ como un campo en $TM$ valorado en $TM$: $\dot
d_{v_a}=v_a$; $\dot d$ es la aplicación idéntica de $TM$ en $TM$.

Una \emph{ecuación diferencial de segundo orden} en $M$ es un
campo tangente $D$ en $TM$ que, como derivación de $\cC^\infty(M)$
en $\cC^\infty(TM)$, coincide con $\dot d$. Entonces, para cada
$v_a\in T_aM$ es $\pi_*(D_{v_a})=v_a$ (siendo $\pi\colon TM\to M$
la proyección canónica). Dos ecuaciones diferenciales de segundo
orden derivan del mismo modo el anillo  $\cC^\infty(M)$, por lo
que su diferencia aniquila a este anillo, luego es un campo
vertical. Las ecuaciones diferenciales de segundo orden en $M$ son
las secciones de un fibrado afín sobre $TM$, modelado sobre el
fibrado vectorial de los campos tangentes verticales.

En el lenguaje de la Física, el valor de una ecuación diferencial
de segundo orden $D$ en el punto $v_a\in TM$ es una
\emph{aceleración} $D_{v_a}$ en la velocidad $v_a$. Un campo
tangente vertical es una \emph{fuerza}. Para que las fuerzas
produzcan aceleraciones es necesaria una estructura adicional en
$M$, como veremos enseguida.

Cada fibra $T_aM$ es un espacio vectorial, por lo que cada vector
$v_a\in T_aM$ es una derivación de $\cC^\infty(T_aM)$, luego de
$\cC^\infty(TM)$, en cada punto $u_a\in T_aM$. Esta derivación
$V_{u_a}$ es un vector vertical, que se llamará
\emph{representante vertical de $v_a$ en $u_a$}; $v_a$ se llamará
\emph{representante geométrico} del vector vertical $V_{u_a}$.
Pasando de vectores en un punto a campos, vemos que cada campo
tangente vertical (\emph{campo de fuerzas}) en $TM$ tiene como
representante geométrico un campo en $TM$ valorado en $TM$, es
decir, una sección de $TM\times_MTM$ sobre el primer factor. En
coordenadas locales, el representante geométrico de
$\left(\frac{\partial}{\partial\dot x^i}\right)_{u_a}$ es
$\left(\frac{\partial}{\partial x^i}\right)_a$, como se ve
acudiendo a las definiciones. Por ejemplo, el generador
infinitesimal del grupo de las homotecias de las fibras de $TM$
es, en coordenadas locales, $\dot
x^i\,\frac{\partial}{\partial\dot x^i}$; su representante
geométrico es $\dot x^i\,\frac{\partial}{\partial x^i}=\dot d$.

Sea $T^*M$ el fibrado cotangente a  $M$; la \emph{forma de
Liouville} $\theta$ en $T^*M$ se define por
$\theta_{\alpha_a}=\alpha_a$ en cada $\alpha_a\in T^*M$. Su
diferencial exterior $\omega_2=d\theta$ es la  \emph{forma
simpléctica} en $T^*M$. En coordenadas locales
$(x^1,\dots,x^n,p_1,\dots,p_n)$ en $T^*M$, se tiene
$$\theta=p_i\,dx^i,\quad\omega_2=d\theta=dp_i\wedge dx^i.$$

Sea $T_2$ una métrica seudoriemanniana (no degenerada) en $M$;
$T_2$ establece un isomorfismo $TM\simeq T^*M$, que hace
corresponder a cada  $v_a\in TM$ la 1-forma $\alpha_a=i_{v_a}T_2$;
diremos que $v_a$ es el \emph{gradiente} de $\alpha_a$; del mismo
modo, cuando el campo tangente $u$ en $M$ y la 1-forma $\alpha$ en
$M$ estén relacionados por $i_uT_2=\alpha$, diremos que $u$ es el
gradiente de $\alpha$. La noción se extiende a campos en $TM$
valorados en $TM$ y 1-formas horizontales en $TM$.

El isomorfismo entre $TM$ y $T^*M$ establecido por $T_2$ nos
permite transportar cualquier estructura de una de estos espacios
al otro. Así, la forma de Liouville $\theta$ de $T^*M$ se
transporta a la 1-forma horizontal en $TM$ que asigna a cada punto
$v_a\in TM$ la forma en $v_a$ que corresponde a
$\theta_{i_{v_a}T_2}=i_{v_a}T_2$, forma que es $i_{v_a}T_2$
(subida por pull-back a $TM$). La forma $\theta$ transportada de
este modo a $TM$ volverá a denotarse por $\theta$; así
$\theta_{v_a}=i_{v_a}T_2$; $\theta_{v_a}$ es el punto de $T^*M$
que le corresponde al punto $v_a$ de $TM$ por el isomorfismo
establecido por la métrica. Cuando $v_a$ varía recorriendo $TM$,
obtenemos $\theta=i_{\dot d}T_2$: el campo tautológico $\dot d$ se
traduce, por la métrica, en la forma tautológica $\theta$.

 En coordenadas locales $(x^i)$ para $M$, $(x^i, \dot x^j)$ para
 $TM$, $(x^i,p_j)$ para $T^*M$, se tienen las expresiones familiares
 $$T_2=g_{ij}dx^idx^j,$$
$p_i=g_{ij}\dot x^j$ para el isomorfismo $TM\simeq T^*M$,
$\theta=g_{ij}\dot x^j\, dx^i$ en $TM$, etc.

La función $T=\frac 12\,\dot\theta$ es la \emph{energía cinética}.

Para evitar que la exposición se recargue con precisiones obvias,
pasaremos de $TM$ a $T^*M$ sin necesidad de advertirlo en cada
caso, sobreentendiendo que usaremos el isomorfismo establecido por
la métrica.
\bigskip

La totalidad de la Mecánica depende del siguiente
\begin{lemacero}
La métrica $T_2$ establece una correspondencia biunívoca entre
ecuaciones diferenciales de segundo orden y 1-formas horizontales
en $TM$, por medio de la ecuación
\begin{equation}\label{lema0}
i_D\omega_2+dT+\alpha=0.
\end{equation}
Los campos tangentes $u$ en $M$ que son integrales intermedias de
la ecuación diferencial de segundo orden $D$ son justamente los
que verifican
\begin{equation}\label{lema01}
i_u\,d(i_uT_2)+dT(u)+u^*\alpha=0,
\end{equation}
donde $u^*\alpha$ es el pull-back de $\alpha$ por la sección
$u\colon M\to TM$ y $T(u)$ es la función $T$ especializada a $u$.
\end{lemacero}

\begin{proof}[Demostración]
(\cite{MecanicaMunoz}, sección 1 y \cite{RAlonso}, pag. 4).

\noindent(\ref{lema0}) Dada la ecuación diferencial de segundo
orden $D$, definimos la forma $\alpha$ por (\ref{lema0}); debemos
comprobar que $\alpha$ es horizontal, es decir, que

$\langle\alpha, V\rangle=0$ para cada campo vertical $V$.

Usando la clásica fórmula de Cartan, tenemos:
$$\langle i_D\omega_2, V \rangle = \langle i_Dd\theta, V \rangle=  D\langle \theta,V \rangle- V   \langle
\theta,D\rangle- \langle \theta, [D,V]\rangle.$$
Como  $\theta$ es horizontal, es   $\langle \theta,V \rangle=0$; por la misma razón,
$\langle \theta,D \rangle=\langle \theta,\dot d \rangle=\dot\theta=2\,T$; y $V\dot\theta=2\langle\theta,v\rangle$, donde
$v$ es el representante geométrico de $V$. Por tanto, tenemos $V\langle\theta, D\rangle=2\,\langle \theta,v \rangle$.

Para $f\in\cC^\infty(M)$ es
$$[D,V]f=-VDf=-V\dot f=-vf,$$
por lo que es $[D,V]=-v+\text{campo vertical}$, luego
$$\langle \theta,[D,V] \rangle =-\langle \theta,v \rangle.$$

Juntando todo, tenemos
 $$\langle i_D\omega_2,V \rangle=0-2\,\langle \theta,v \rangle+\langle
   \theta,v \rangle=-\langle \theta,v \rangle.$$
   Además,
   $$\langle dT,V\rangle=VT=\frac 12\,V\dot\theta=\langle
   \theta,v   \rangle.$$
   Sumando, queda $\langle \alpha,V \rangle =0$: $\alpha$ es
   horizontal.

   Como $\omega_2$ no tiene radical, la correspondencia
   $D\to\alpha$ es inyectiva. Por otra parte, si $V$ es cualquier
   campo vertical, $D+V$ es una ecuación de segundo orden y, de lo
   ya demostrado resulta que $i_V\omega_2$ es horizontal. Como los
   campos verticales y las 1-formas horizontales son
   $\cC^\infty(TM)$-módulos  del mismo rango y localmente
   libres, se deduce que, dada cualquier $\alpha$ horizontal,
   existe un $D$ que cumple (\ref{lema0}).
   \bigskip

   \noindent(\ref{lema01}) Pensando $u$ como sección de $TM\to M$, el que
   sea integral intermedia de $D$ significa que $D$ es tangente a
   $u$. La especialización de $\theta$ a la sección $u$ es
   $i_uT_2$ (subida a la sección $u$), luego la de $\omega_2$ es
   $di_uT_2$. Así, la condición (\ref{lema01}) sobre $u$ significa
   que la especialización de la 1-forma $i_{u_*(u)}\omega_2$ a la
   sección $u$ es la de $-dT(u)-\alpha$ que, por (\ref{lema0}), es
   la especialización de $i_D\omega_2$. Por tanto, la condición
   (\ref{lema01}) significa que el campo vertical $V=D-u_*u$ (con
   soporte en $u$) verifica $i_V\omega_2|_u=0$, y como
   $i_V\omega_2$ es horizontal,  esto equivale a que sea $V=0$.

   \end{proof}

   \begin{obs*}
   En la demostración de (\ref{lema0}) hemos visto que, para cada
   campo vertical $V$, la 1-forma $i_V\omega_2$ es horizontal.
   Esto puede verse inmediatamente usando coordenadas de $T^*M$;
   los campos verticales de $TM$ pasan a campos verticales
   (tangentes a las fibras) en $T^*M$, combinaciones lineales de
   los $\frac{\partial}{\partial p_i}$; como es
   $\omega_2=dp_i\wedge dx^i$, se concluye. De modo más preciso,
   se tiene
   \begin{equation}\label{notalema0}
   i_V\omega_2=i_vT_2
   \end{equation}
   si $v$ es el representante geométrico del campo vertical $V$.
   La demostración en coordenadas es fácil.

   \end{obs*}

Un \emph{sistema mecánico clásico} es una variedad $M$ (el
\emph{espacio de configuración}) provista de una métrica
seudoriemanniana $T_2$ y una 1-forma horizontal $\alpha$, la
\emph{forma de trabajo} o \emph{forma de fuerza}. La  ecuación
diferencial de segundo orden $D$ que corresponde a  $\alpha$ por
 (\ref{lema0}) es  la \emph{ecuación diferencial del movimiento} del
 sistema $(M,T_2,\alpha)$ y (\ref{lema0}) es la \emph{ecuación de Newton}.

Para $\alpha=0$ se tiene el sistema \emph{libre de fuerzas} o
\emph{geodésico}, cuya ecuación del movimiento es el \emph{campo
geodésico}, que denotaremos $D_G$: $i_{D_G}\omega_2+dT=0$.

El campo geodésico proporciona el origen para el fibrado afín de
las ecuaciones diferenciales de segundo orden. Para cada ecuación
diferencial de segundo orden $D$, $D-D_G=V$ es un campo vertical,
la \emph{fuerza} del sistema $(M,T_2,\alpha)$ cuya ecuación del
movimiento es $D$. El representante geométrico de la fuerza $V$ es
un campo en  $TM$ con valores en $TM$, al que denotaremos por
$D^\nabla$ y llamaremos \emph{valor covariante} de $D$.

 Restando las ecuaciones (\ref{lema0}) que corresponden a $D$ y $D_G$ resulta  $i_V\omega_2+\alpha=0$, y aplicando
(\ref{notalema0}):
\begin{equation}\label{Newton}
i_{D^\nabla}T_2+\alpha=0\quad\text{ó}\quad
D^\nabla=-\textrm{grad}\,\alpha
\end{equation}
 Esta es la versión general de la ley de Newton
 ``fuer\-za = ma\-sa $\cdot$ ace\-le\-ra\-ción''; aquí la fuerza es
  $-\textrm{grad}\,\alpha$ (representante geométrico de un campo
  vertical) y ``ma\-sa$\cdot$a\-ce\-le\-ra\-ción'' es $D^\nabla$, cuando se
  entiende que la ``masa'' está incorporada en los coeficientes de
  la métrica (Lagrange) y se observa que, para cada curva en $M$
  que sea solución de la ecuación diferencial de segundo orden $D$,
  el vector tangente $u$ a lo largo de la curva verifica $u^\nabla
  u=D^\nabla$ (ver fórmula (22) en \cite{MecanicaMunoz}). Para
  referencias posteriores, expresamos en coordenadas la
  (\ref{Newton}) para $\alpha=A_i\,dx^i$:
  \begin{equation}\label{Newtoncoordenadas}
  g_{lk}\ddot x^l+\Gamma_{ij,k}\dot x^i\dot x^j+A_k=0
  \end{equation}
  (los detalles del cálculo, en \cite{MecanicaMunoz}, sección
  1.10).

En los ``Principia'' demostró Newton que el cumplimiento de las tres leyes de Kepler para \emph{cualquier} partícula que se mueva alrededor del
Sol implica la Ley de la Gravitación Universal que lleva su nombre. El resultado general en esta cuestión es una consecuencia inmediata del Lema 0:
\begin{thm*}
Dado el espacio de configuración $(M,T_2)$, para cada campo tangente $u$ en $M$ existe una única forma de fuerza $\alpha$ que sea independiente de la
velocidad y haga que $u$ sea una integral intermedia del sistema mecánico $(M,T_2,\alpha)$.
\end{thm*}
\begin{proof}[Demostración]
Por (\ref{lema01}) y por ser $\alpha$ independiente de las velocidades, es
$$\alpha=u^*\alpha=-i_ud\left(i_uT_2\right)-dT(u)$$
\end{proof}

 Un \emph{sistema de ligaduras} en $(M,T_2,\alpha)$ es un sistema
 de Pfaff $\Lambda$ en $TM$ consistente en formas horizontales;
 cuando $\Lambda$ esté generado por un sistema de Pfaff en $M$, se
 dice que el sistema es de \emph{ligaduras lineales}. En el
 sistema con ligaduras $(M,T_2, \alpha,\Lambda)$ la ecuación de
 Newton
(\ref{lema0}) se sustituye por la congruencia
\begin{equation}
\label{Newtonligada}
i_D\omega_2+dT+\alpha\equiv 0\,\, (\textrm{mod}\,\Lambda)
\end{equation}
que se impone al campo $D$, junto con el \emph{Principio de los
Trabajos Virtuales}: las trayectorias permitidas son soluciones de
$D$ que (en $TM$) quedan dentro del conjunto definido por las
ecuaciones $\dot\beta=0$ ($\forall\beta\in\Lambda$).

La determinación de  $D$ por  (\ref{Newtonligada}) y el Principio
de los Trabajos Virtuales es posible en condiciones adecuadas
sobre la métrica y las ligaduras: por ejemplo, en el caso más
clásico de métrica definida positiva y ligaduras lineales. Ver
\cite{MecanicaMunoz}, sección~3.
\medskip

Terminaremos esta sección con algunas observaciones sobre el
llamado \emph{Principio de Reversibilidad Microscópica} (ver
\cite{Sommerfeld}, por ejemplo), según el cual todos los procesos
fundamentales en el Mundo Físico son reversibles, siendo la
irreversibilidad un efecto estadístico-macroscópico, en el caso
``clásico'' y, en el caso ``cuántico'', además, del colapso de la
función de onda en el proceso de medida.

Limitándonos a nuestro tema, los sistemas mecánico-clásicos, la
re\-ver\-si\-bi\-li\-dad de un sistema $(M,T_2,\alpha)$ significa
que una curva-solución  (en $M$) de la ecuación di\-fe\-ren\-cial de
segundo orden $D$ que rige la evolución, al ser parametrizada
cambiando el signo del parámetro, es decir, al ser recorrida en
sentido inverso, se transforma en otra solución de $D$. En
general, esto no ocurre. Hay sistemas mecánico-clásicos
irreversibles. Sin entretenernos en una discusión general, se ve
en (\ref{Newtoncoordenadas}) que, si los coeficientes $A_k$ de la
forma de trabajo son funciones homogéneas de grado $r$ de las
$\dot x$, cuando $r$ es par el sistema es reversible (por ejemplo,
en fuerzas de tipo Coulomb) y cuando $r$ es impar, el sistema es
irreversible (por ejemplo, fuerzas de tipo Lorentz). En el caso
del electromagnetismo se arregla la irreversibilidad imponiendo a
las ``partículas de prueba'' que sigan las trayectorias de la
corriente de Maxwell, con lo que el recorrido a la inversa solo es
posible si se cambia antes el signo del campo electromagnético, y
con ello la forma de fuerza $\alpha$. Dejamos esto aquí.

\section{Sistemas conservativos}\label{sconservativos}

Cuando $\alpha$ es una diferencial exacta, el sistema
$(M,T_2,\alpha)$ se dice que es  \emph{conservativo}. Como
$\alpha$ es horizontal, la función potencial $U$ de la que
$\alpha$ es diferencial pertenece a $\cC^\infty(M)$. La función
suma $H=T+U$ se llama \emph{hamiltoniano} del sistema. La ecuación
de Newton (\ref{lema0}) es, en este caso:
\begin{equation}\label{newtonconservada}
i_D\omega_2+dH=0.
\end{equation}
Escrita en coordenadas $(x,p)$ de $T^*M$, (\ref{newtonconservada})
es el sistema de \emph{ecuaciones canónicas de Hamilton}. En cada
hipersuperficie $H=\textrm{cte.}$ de $T^*M$ la especialización de
$\omega_2$ tiene como radical $D$ (y sus múltiplos), como se
deduce de (\ref{newtonconservada}); del argumento clásico basado
en el teorema de Stokes resulta, entonces, el \emph{Principio de
Maupertuis}: Las curvas en $M$ con extremos dados y parametrización
tal que, subidas a $TM$ queden dentro de una misma hipersuperficie
$H=\textrm{cte.}$, dan valores a $\int\theta$ que son extremales
justamente para las trayectorias del sistema.

La ecuación (\ref{lema01}) para las integrales intermedias de $D$
es, en este caso:
\begin{equation}\label{intermediaconservada}
i_ud(i_uT_2)+dH(u)=0
\end{equation}
En particular, si $u$ es una subvariedad lagrangiana de $TM$, es
$d(i_uT_2)=d\theta|_u=0$, y la ecuación de Newton es
$$dH(u)=0\quad\text{ó}\quad H(u)=\textrm{cte.}$$
Esta es la \emph{ecuación de Hamilton-Jacobi} cuando se traduce a
la función $S$ de la que $u$ es gradiente:
\begin{equation}\label{HamiltonJacobi}
H(\textrm{grad}\,S)=\textrm{cte.}\quad\text{en $TM$,}\qquad
H(dS)=\textrm{cte.}\quad\text{en $T^*M$}
\end{equation}
Es ésta una ecuación en derivadas parciales de primer orden para
la función $S$ que recoge toda la estructura del sistema mecánico;
pero dejamos de lado este tema.
Señalemos solamente una consecuencia del teorema de la sección 0 y el teorema de existencia de integrales completas para las ecuaciones en derivadas parciales de primer orden:
\begin{prop*}
Un sistema mecánico $(M,T_2,\alpha)$ en que la forma de trabajo $\alpha$ es independiente de las velocidades, es conservativo si y solo si admite una integral intermedia lagrangiana (con más rigor: una familia de integrales intermedias lagrangianas locales cuyos dominios de existencia cubran $M$).
\end{prop*}

Recordemos que un campo tangente $u$ en $M$ se llama
\emph{conservativo} cuando $\textrm{div}\,u=0$. Cuando $u$ es una
sección lagrangiana de $TM$, es $u=\textrm{grad}\,S$, para una
función $S$, y la condición de conservativo es $\Delta S=0$.
\begin{thm*}
Sea $(M,T_2,dU)$ un sistema mecánico conservativo. Sea $S\in
\cC^\infty(M)$. De las tres condiciones
\begin{enumerate}[A)]
\item  $S$ verifica la ecuación de Hamilton-Jacobi
(\ref{HamiltonJacobi}): $$H(\textrm{grad}\,
S)=E,\quad\text{constante},$$
\item $S$ es armónica: $\Delta S=0$,
\item $\Psi=e^{iS}$ verifica la ecuación de Schrödinger:
$$\left(-\frac 12\Delta+U\right)\Psi=E\Psi,$$
\end{enumerate}
cada par de ellas implica la tercera.
\end{thm*}
\begin{proof}[Demostración]
Calculamos $\Delta\Psi$:
$$\textrm{grad}\,\Psi=i\Psi\,\textrm{grad}\,S;$$
\begin{align*}
\Delta\Psi&=\textrm{div}\,\textrm{grad}\,\Psi
            =i\textrm{div}(\Psi\textrm{grad}\,S)=i\Psi\textrm{div}\,\textrm{grad}\,S+i\textrm{grad}\,S(\Psi)\\
         &=i\Psi\Delta S-\Psi\textrm{grad}\,S(S)=
           [i\Delta
           S-T_2(\textrm{grad}\,S,\textrm{grad}\,S)]\Psi\\
         &=[i\Delta S-2(H(\textrm{grad}\,S)-U)]\Psi.
\end{align*}
Queda:
\begin{equation}\label{Schrodinger}
\left[-\frac 12\Delta+U\right]\Psi=
   \left[\frac 1{2i}\Delta S+H(\textrm{grad}\,S)\right]\Psi.
\end{equation}
De esta identidad se deduce el teorema.
\end{proof}

\begin{obs*}
Las dos primeras condiciones del teorema caracterizan a los campos
tangentes a $M$ que son integrales intermedias del sistema
mecánico y son lagrangianos ($u=\textrm{grad}\,S$) y conservativos
($\textrm{div}\,u=0$).
\end{obs*}

Para terminar esta sección vamos a mostrar la relación entre
sistemas libres de fuerzas y sistemas conservativos en dimensión
una unidad menor.

Sea $M$ de dimensión $n=m+1$, con métrica $T_2$. Sea $x^0$ una
función en $M$, cuya diferencial no se anula en ningún punto.
Tomamos como coordenadas locales en $M$ la $x^0$ junto con $m$
integrales primeras $x^\mu$ de $\textrm{grad}\,x^0$. Con estas
coordenadas, la métrica toma la forma
\begin{equation}\label{metricatiempo}
T_2=g_{00}(dx^0)^2+g_{\mu\nu}dx^\mu\,dx^\nu\quad \mu,\nu=1,\dots,m.
\end{equation}

Supongamos que $T_2$ es proyectable al anillo de integrales
primeras de $\textrm{grad}\,x^0$; es decir, que, si $f$, $g$ son
integrales primeras de $\textrm{grad}\,x^0$, también lo es
$T^2(df,dg)$; con las coordenadas que usamos, esta condición
significa que las $g_{\mu\nu}$ no dependen de $x^0$. Finalmente,
supongamos que $g_{00}$ tampoco depende de $x^0$; esta condición
 puede expresarse intrínsecamente por
  $\textrm{II}_{\textrm{grad}\,x^0}(\textrm{grad}\,x^0,\textrm{grad}\,x^0)=0$
  ($\textrm{II}_{\textrm{grad}\,x^0}$ es la segunda forma fundamental del
  campo $\textrm{grad}\,x^0$). Con estas condiciones, los símbolos
  de Christoffel $\Gamma_{ij,k}$ con un solo índice 0 ó los tres
  0, son nulos. Las ecuaciones (\ref{Newtoncoordenadas}) para el
  campo geodésico ($A_k=0$) quedan
  \begin{equation}\label{conservativo}
  \begin{cases}
  g_{00}\ddot x^0+2\Gamma_{0\mu,0}\dot x^0\dot x^\mu=0 &\\
  g_{\mu\nu}\ddot x^\nu+\Gamma_{\sigma\rho,\mu}\dot
  x^\sigma\dot x^\rho+\Gamma_{00,\mu}(\dot x^0)^2=0 &(\mu=1,\dots,m)
  \end{cases}
  \end{equation}
 Sustituyendo los símbolos de Christoffel por sus valores:
 $$g_{00}\ddot x^0+\frac{\partial g_{00}}{\partial x^\mu}\dot x^0\dot
 x^\mu=0,\quad\text{que es}\quad D_G(g_{00}\dot x^0)=0$$
 luego $g_{00}\dot x^0$ es una integral primera de $D_G$;
 denotémosla $E_0/c$.

 Las ecuaciones para los índices $\mu\ne 0$ son:
 $$g_{\mu\nu}\ddot x^\nu+\Gamma_{\sigma\rho,\mu}\dot
  x^\sigma\dot x^\rho-\frac 12 (\dot x^0)^2\frac{\partial g_{00}}{\partial
  x^\mu}=0,$$
  luego
 \begin{equation}\label{conservativo2}
  g_{\mu\nu}\ddot x^\nu+\Gamma_{\sigma\rho,\mu}\dot
  x^\sigma\dot x^\rho+\frac {E_0^2}{2c^2} \frac{\partial g^{00}}{\partial
  x^\mu}=0,\quad\text{(en $g_{00}\dot x^0=E_0/c$)}
\end{equation}
 Comparando con (\ref{Newtoncoordenadas}) vemos que
 (\ref{conservativo2}) son las ecuaciones para un sistema
 conservativo en $M'$ (proyección de $M$ por
 $\textrm{grad}\,x^0$), con métrica $T'_2$ (proyección de $T_2$) y
 energía potencial
 \begin{equation}\label{conservativo3}
 U=\frac {E_0^2}{2c^2}g^{00}+\textrm{cte}\quad \text{($\alpha=dV$)}
 \end{equation}
 A la inversa, es claro que todo sistema conservativo en dimensión
 $m$ procede, por proyección, de un sistema geodésico de dimensión
 $m+1$.

 Por ejemplo, $\mathbb{R}^3$ con potencial newtoniano
 $\displaystyle{-\frac{\textrm{cte}}r}$ ($\textrm{cte}>0$) se obtiene por
 pro\-yec\-ción de una especialización del sistema geodésico en
 $\mathbb{R}^4$ con una métrica de signatura Minkowski, etc.

 Esta especie de transferencia de energía cinética en $M$ a energía potencial en $M'$ podría tener relación con antiguas ideas de Hertz
 \cite{Gantmajer}, VII; pero no hemos estudiado la cuestión.

 La ecuación de Hamilton-Jacobi para el sistema geodésico $(M,T_2,0)$ (ecuación de la ``eikonal'') es
 $$g^{00}\left(\frac{\partial S}{\partial x^0}\right)^2+
          g^{\mu\nu}\frac{\partial S}{\partial x^\mu}\frac{\partial S}{\partial x^\nu}=2W\quad\text{constante}.$$

 En esta ecuación puede separarse la variable $x^0$ de las demás poniendo
 $$\frac{\partial S}{\partial x^0}=\pm\frac{E_0}{c}$$
luego
$$S=\pm E_0\frac{x^0}c+S'(x^1,\dots,x^m),$$
donde $S'$ satisface la ecuación
$$g^{00}\frac{E_0^2}{c^2}+g^{\mu\nu}\frac{\partial S'}{\partial x^\mu}\frac{\partial S'}{\partial x^\nu}=2W,$$
que es la ecuación de Hamilton-Jacobi para el sistema conservativo $(M',T'_2, dV)$
(proyección del sistema geodésico $(M,T_2,0)$) en el que el potencial es
$V=\frac 12\frac{E_0^2}{c^2}g^{00}$, como acabamos de ver.

La acción en el sistema geodésico en $M$ es
\begin{equation}\label{acciontiempo}
S=\pm E_0\frac{x^0}c+S',
\end{equation}
la acción ``con tiempo'' (en el término $E_0t$, $t=x^0/c$), que no es la acción $S'$ en el sistema conservativo proyectado.

La separación de variables que se da en este caso en la ecuación de Hamilton-Jacobi es la aplicación canónica
del método de Jacobi para las ecuaciones en derivadas parciales de primer orden:
 el campo geodésico es el campo hamiltoniano correspondiente a $T=\frac 12\dot\theta$; su integral primera $p_0$ conmuta con $T$
 (para el paréntesis de Poisson); el método de Jacobi se aplica, entonces, añadiendo a la ecuación $T(dS)=W$ la ecuación
 $p_0(dS)=\textrm{cte}$.

\section{Tiempo. Ligaduras de tiempo}

En Mecánica Clásica, el tiempo es el parámetro de cada
curva-solución del campo $D$ que rige la evolución del sistema. No
existe ninguna función en el anillo $\cC^\infty(M)$ que pueda
parametrizar todas las curvas-solución de una ecuación diferencial
de segundo orden $D$ (tal función $f$ verificaría $Df(=\dot f)=1$
idénticamente, lo que es absurdo). Tampoco existe función en el
anillo $\cC^\infty(TM)$ que parametrice a todas las curvas
solución de todas las ecuaciones diferenciales de segundo orden,
porque tal función sería aniquilada por los campos verticales,
luego estaría en $\cC^\infty(M)$.

El objeto natural que parametriza a todas las curvas-solución de
todas las ecuaciones diferenciales de segundo orden en $M$ es una
clase de 1-formas en $TM$, a la que llamaremos \emph{clase del
tiempo}, constituida por las 1-formas  horizontales
 $\alpha$ tales que $\dot\alpha=1$. Para cada 1-forma horizontal
 $\beta$, en el abierto en que es $\dot\beta\ne 0$, la forma
 $\beta/\dot\beta$ está
 en la clase del tiempo. Dos formas de la clase del tiempo difieren, en su dominio común de definición,
 en una forma del \emph{sistema de contacto}  $\Omega$ de $TM$;
 recordemos que $\Omega$ es el sistema de Pfaff en $TM$ incidente
 con todas las ecuaciones diferenciales de segundo orden; en cada
 abierto coordenado, $\Omega$ está generado por las formas
 $\dot x^idx^j-\dot x^jdx^i$.

 En el abierto de $TM$ en que $\dot\theta\ne 0$, la forma $\theta/\dot\theta$ es un representante
 de la clase del tiempo.

 Podemos  \emph{elegir un tiempo} para $M$ eligiendo una forma horizontal  $\tau$ en $TM$
 y admitiendo como posibles estados de posición-velocidad solo los puntos de $TM$
 que verifiquen la ecuación $\dot\tau=1$. Es lo que llamaremos
 \emph{ligadura de tiempo}. En particular, para $\tau=dt$, con $t$
 una función local en $M$, la ligadura de tiempo $\dot t=1$
 selecciona las trayectorias de la ecuaciones diferenciales de
 segundo orden en que la función $t$ ``fluye uniformemente'' (como
 el ``tiempo absoluto'' de Newton).

 La imposición de una ligadura de tiempo $\tau$ a un sistema
 mecánico dado $(M,T_2,\alpha)$, modifica el campo $D$ que rige
 el movimiento antes de la ligadura, a otro campo $\overline
 D$, de modo análogo a las ligaduras ordinarias. Lo natural es imponer que las
 ligaduras ordinarias sean un límite de ligaduras de tiempo,
 pasando de $\dot\tau=1$ a $\dot\tau=0$ a través de las ligaduras
 de tiempo $\dot\tau=\textrm{cte.}$ $D$ se modifica en el campo
  $\overline D$ que verifica la congruencia
 \begin{equation}\label{Newtontiempo}
i_{\overline D}\omega_2+dT+\alpha\equiv 0\,\, (\textrm{mod}\,\tau)
\end{equation}
y, en sustitución del Principio de los Trabajos Virtuales:
$$\overline D(\dot\tau)=0,\quad\text{($\overline D$ es tangente a las variedades $\dot\tau=\textrm{cte.}$)}$$

El campo que satisface estas condiciones, en el abierto de
$TM$  en que $\|\textrm{grad}\,\tau\|$ no se anula, es
\begin{equation}\label{Campotiempo1}
\overline D=D-\frac {D\dot\tau}{\|\tau\|^2}
  \,\textrm{Grad}\,\tau
\end{equation}
donde $\textrm{Grad}\,\tau$ es el campo vertical (fuerza) cuyo
representante geométrico es $\textrm{grad}\,\tau$. En
\cite{MecanicaMunoz}, sección 3.2, se demuestra que, cuando $\tau$
es una 1-forma de $M$, (\ref{Campotiempo1}) equivale a
\begin{equation}\label{Campotiempo2}
\overline D=D-\frac 1{\|\tau\|^2}
  \left(\langle\tau,D^\nabla\rangle+\textrm{II}_{\textrm{grad}\,\tau}(\dot
  d,\dot d)\right)\,\textrm{Grad}\,\tau
\end{equation}
donde $\textrm{II}_u$ es la segunda forma fundamental del campo $u$
respecto de $T_2$.

Cuando $D$ es el campo geodésico queda solo el segundo término de
los que modifican $D$.

Vimos en el apartado 1 que, en las condiciones allí precisadas, el
campo geodésico de una variedad $M$  de dimensión $m+1$ se
especializa a ciertas subvariedades de  $TM$ y, de ellas, se
proyecta como campo de un sistema conservativo en una $M'$ de
dimensión $m$. Volvamos a aquellas notaciones, con las ecuaciones
(\ref{conservativo}) para las geodésicas de $M$. Impongamos al
campo geodésico $D_G$ la ligadura de tiempo $\dot x^0=1$; en la
congruencia (\ref{Newtontiempo}) es $\tau=dx^0$; los coeficientes
de las $dx^\mu$ ($\mu\ne 0$) en el primer miembro de
(\ref{Newtontiempo}) deben ser los mismos con $\overline D$ que
eran con $D$. Por ello, la modificación de la segunda fila en
(\ref{conservativo}) por la ligadura $\dot x^0=1$ da
\begin{equation}\label{Campotiempo3}
  g_{\mu\nu}\ddot x^\nu+\Gamma_{\sigma\rho,\mu}\dot
  x^\sigma\dot x^\rho-\frac 12 \frac{\partial g_{00}}{\partial
  x^\mu}=0.
\end{equation}
Estas son ecuaciones del movimiento para un sistema conservativo
$(M', T'_2, dU)$, donde la energía potencial es ahora $-g_{00}/2$.
Obsérvese el cambio de $g^{00}$ por $-g_{00}$ de
(\ref{conservativo2}) a (\ref{Campotiempo3}).

Podemos enunciar:
\begin{thm*}
Todo sistema mecánico conservativo es una ligadura de tiempo
holónoma ($\tau$ es una diferencial exacta $dx^0$) en un sistema
libre. Y también es la proyección de la especialización de un
sistema libre a una ligadura no holónoma de tiempo
($\tau=g_{00}dx^0$) a la que es tangente el campo geodésico.
\end{thm*}

De modo general, consideremos un sistema conservativo $(M,T_2,dU)$ y \emph{elijamos como tiempo} una función $x^0$
(limitando $M$ al abierto en que $dx^0$ no es cero). Si el campo que rige la evolución del sistema dado es $D$ (el campo hamiltoniano
correspondiente a $H=T+U$), la ligadura de tiempo $\dot x^0=1$ lo modifica en el campo $\overline D$ dado por (\ref{Campotiempo1}).
Tomando las coordenadas locales $x^0,x^1,\dots,x^m$ de modo que las $x^\mu$ (con $\mu=1,\dots,m$) sean integrales primeras de
$\textrm{grad}\,{x^0}$,
la métrica toma la forma (\ref{metricatiempo}); ahora no suponemos nada sobre la dependencia de las $g$ respecto de $x^0$.

Hechos los cálculos a partir de (\ref{Campotiempo1}) se encuentra, en la variedad $\dot x^0=1$:
\begin{equation}
\overline D|_{\dot x^0=1}=\frac{\partial }{\partial  x^0}
                               +\dot g_{00}\frac{\partial }{\partial  p_0}
                               +\frac{\partial h}{\partial  p^\mu}\frac{\partial }{\partial  x^\mu}
                               -\frac{\partial \left(h-\frac{g_{00}}2\right)}{\partial  x^\mu}\frac{\partial }{\partial  p_\mu},
\end{equation}
donde es
$$h=\frac 12 g^{\mu\nu}p_\mu p_\nu+U\quad\left( =H-\frac 12 g^{00}p_0^2 \right).$$

El sistema de ecuaciones diferenciales que rige la evolución del sistema en la ligadura $\dot x^0=1$ es
\begin{equation}\label{canonicastiempo}
\begin{cases}
\displaystyle{\,\,\frac{dx^0}{dt}=1},                 &\displaystyle{\frac{dp_0}{dt}=\dot g_{00}}\\
 &\\
\displaystyle{\,\,\frac{dx^\mu}{dt}=\frac{\partial H'}{\partial  p_\mu}}, \,\,
                              & \displaystyle{\frac{dp_\mu}{dt}=-\frac{\partial H'}{\partial  x^\mu}},
                              \quad\text{con}\quad H'=h-\frac 12 g_{00}=H|_{\dot x^0=1}-g_{00}
 \end{cases}
  \end{equation}
\bigskip

La ecuación de Hamilton-Jacobi para sistemas ``disipativos'', en que el hamiltoniano depende del ``tiempo'', se deduce habitualmente dando un rodeo por el principio variacional de Hamilton (del que trataremos en la sección 5), tomando como dato fundamental una lagrangiana, a partir de la que se definen los momentos, el hamiltoniano, y la acción; así, por ejemplo, en (\cite{SommerfeldMecanica}, VIII, 44(2)). Si queremos mantener nuestro punto de vista, no podemos seguir ese camino. Debemos llegar a la ecuación de Hamilton-Jacobi como caracterización de ciertas integrales intermedias de las ecuaciones del movimiento, canónicamente asociadas al sistema hamiltoniano y la ligadura de tiempo impuesta.

Partimos de un sistema conservativo $(M,T_2,dU)$ en el  que se impone la ligadura de tiempo $\dot x^0=1$. El campo $D$ que rige la evolución del sistema antes de la ligadura (antes de la imposición de $x^0$ como tiempo) se modifica según (\ref{Campotiempo1}) en $\overline D$, con $\tau=dx^0$, en nuestro caso. Manteniendo las notaciones anteriores, las ecuaciones del movimiento son ahora las (\ref{canonicastiempo}); el segundo grupo de estas ecuaciones corresponde a un sistema ``disipativo''  en $M'$, con hamiltoniano $H'$ dependiente del ``tiempo'' $x^0$. Este hamiltoniano $H'$ no es directamente el resultado de sustituir $\dot x^0$ por 1 en $H$, salvo el caso en que el coeficiente $g_{00}$ de la métrica en $M$ sea independiente de las coordenadas ``espaciales'' $x^\mu$; en cuyo caso puede mantenerse $H$ sin más que sustituir $\dot x^0$ por 1.

Para llegar a la ecuación de Hamilton-Jacobi en este caso, vamos a partir de una variante del Lema 0;
en primer lugar, expresamos (\ref{Newtontiempo})  como ecuación de Newton para el campo $\overline D$:
 \begin{equation}\label{Newtontiempo1}
 i_{\overline D}d\theta+dH+A\,dx^0=0;
 \end{equation}
  contrayendo con $\overline D$, resulta $\overline DH+A\,\overline Dx^0=0$, luego $A=-\overline DH/\dot x^0$. Sustituyendo y agrupando podemos reescribir (\ref{Newtontiempo1}) como
 \begin{equation}\label{Newtontiempo2}
 i_{\overline D}\left(d\theta-dH\wedge\frac{dx^0}{\dot x^0}\right)=0.
 \end{equation}
 Por otra parte, modificando la forma de Liouville $\theta$ como
 \begin{equation}\label{Liouvilletiempo}
 \overline\theta:=\theta-H\frac{dx^0}{\dot x^0}=\left(p_0-\frac H{\dot x^0}\right)dx^0+p_1dx^1+\cdots+p_mdx^m
 \end{equation}
 resulta una nueva estructura simpléctica sobre $TM$,
 $$d\overline\theta=d\theta-dH\wedge\frac{dx^0}{\dot x^0}+H\frac{d\dot x^0}{(\dot x^0)^2}\wedge dx^0.$$
 Utilizando $\overline D\dot x^0=0$ y $\overline Dx^0=\dot x^0$, (\ref{Newtontiempo2}) se reformula como la ecuación  (\ref{Newtontiempo3})
 del siguiente enunciado, el resto del cual se demuestra con argumentos similares a los utilizados en el Lema 0.
 \begin{lema*}\label{lema0tiempo}
 Sean $D$ la ecuación de segundo orden correspondiente al sistema mecánico $(M,T_2,dU)$, $x^0$ una función en $M$ y $\overline\theta=\theta-\frac{H}{\dot x^0}dx^0$. El campo
  $\overline D$ que resulta de modificar $D$ por la ligadura de tiempo $\dot x^0=\textrm{cte}$, está caracterizado por
 \begin{equation}\label{Newtontiempo3}
 i_{\overline D}d\overline\theta+\frac H{\dot x^0}\,d\dot x^0=0.
 \end{equation}

 Los campos tangentes $u$ en $M$ que son integrales intermedias  de $\overline D$ son aquellos tales que
 \begin{equation}\label{intermediatiempo}
 i_ud(u^*\overline\theta)+\frac{H(u)}{u^0}\,du^0=0,\quad \text{donde }\, u^0:=u(x^0)=u^*\dot x^0.
 \end{equation}
 \end{lema*}

 En particular, son integrales intermedias todos los campos definidos por sub\-va\-rie\-da\-des lagrangianas para la forma simpléctica  $d\overline\theta$ (es decir, tales  que $u^*d\overline\theta=0$), a condición de que
 $u^0=u^*\dot x^0$ sea constante; en el caso en el que dicha constante es 1, se tiene localmente:
 $$ u^*\overline\theta=i_uT_2-H(u)\,dx^0=dW,$$
 siendo $W$ una función de $M$.

Si, como hemos hecho antes, incluimos $x^0$ en un sistema de coordenadas $x^0$, $x^1$, $\dots$, $x^m$, con $T^2(dx^0,dx^\mu)=0$, $\mu=1,\dots,m$, y ponemos
$i_uT_2=p_0(u)\,dx^0+p_1(u)\,dx^1+\cdots +p_m(u)\,dx^m$, la condición anterior se escribe:
$$(p_0-H)(u)=\frac{\partial W}{\partial x^0},\quad p_\mu(u)=\frac{\partial W}{\partial x^\mu}$$
para una cierta función $W$ en $M$; es decir,
$$\frac{\partial W}{\partial x^0}+H\left(x^0,x^1,\dots,x^m,p_0(u), \frac{\partial W}{\partial x^1},\dots,\frac{\partial W}{\partial x^m}\right)=
                                   p_0(u),$$
que junto con la ligadura $p_0(u)=g_{00}u^*\dot x^0=g_{00}$,  dá:
\begin{equation}\label{HJtiempo}
\frac{\partial W}{\partial x^0}+H'\left(x^0,x^1,\dots,x^m, \frac{\partial W}{\partial x^1},\dots,\frac{\partial W}{\partial x^m}\right)=0
\end{equation}
que es la forma analítica de la ecuación de Hamilton-Jacobi para nuestro caso (ecuación en derivadas parciales de la integrales intermedias $d\overline\theta$-lagrangianas). Ob\-sér\-ve\-se que es
$$H'=\frac 12 g^{\mu\nu}p_\mu p_\nu+U-\frac 12 g_{00}.$$

Cuando las $g_{ij}$ y $U$ son independientes de $x^0$, se separan variables en (\ref{HJtiempo}) poniendo
\begin{equation}\label{HJseparadatiempo}
W=S(x^1,\dots,x^m)-E x^0,\quad\text{$E$ constante,}
\end{equation}
donde $S$ es solución de la ecuación de Hamilton-Jacobi para el sistema conservativo $(M',T'_2, d(U-g_{00}/2))$:
\begin{equation}\label{HJtiempo2}
\frac 12\, g^{\mu\nu}\frac{\partial S}{\partial x^\mu}\frac{\partial S}{\partial x^\nu}+U-\frac 12 g_{00}=E
\end{equation}

Para $U=0$ queda la ecuación que corresponde al sistema conservativo en $M'$ producido por una ligadura de tiempo en el sistema geodésico en $M$. Aunque (\ref{HJseparadatiempo}) tenga la apariencia de (\ref{acciontiempo}) la diferencia está en que en (\ref{HJseparadatiempo}) $x^0$ es un verdadero tiempo, que parametriza las trayectorias, mientras que en (\ref{acciontiempo}) no lo es, en general; allí es una coordenada ``tiempo'' como la del espacio de Minkowski de la Relatividad.

En el caso en que las $g_{ij}$ y la $U$ sean independientes de $x^0$ se demuestra igual que en el caso sin ligadura de tiempo que, si la acción (\ref{HJseparadatiempo}) es armónica en $(M,T_2)$, la función $\Phi=e^{iW}$ verifica la ecuación de Schrödinger con tiempo $t=x^0$:
\begin{equation}\label{Schtiempo}
\left(-\frac 12 \Delta'+U-\frac{g_{00}}2\right)\Phi=i\frac{\partial\Phi}{\partial t},
\end{equation}
donde $\Delta'$ es la laplaciana en el ``espacio'' $(M',T'_2)$.
\medskip

\section{Tiempo propio. Fuerzas relativistas}

De los tiempos heroicos en que Einstein derrotó al Éter ha
llegado hasta los nuestros, como eco de ecos, una confusa
distinción entre ``relativista'' y ``no relativista''.

No nos detendremos en señalar errores que aparecen en los textos
como productos de esa confusión. Vamos directamente a nuestra
cuestión: ¿\emph{qué sig\-ni\-fica que un sistema mecánico}
$(M,T_2,\alpha)$ \emph{es relativista o no lo es}? Sabemos que en
la Mecánica Lagrangiana no hay una función ``tiempo'', salvo que
se imponga (como ligadura, por ejemplo) de modo que, desde el punto de vista
general en que nos situamos, decir que ``el tiempo es una
coordenada como las otras'' es vacío.

La signatura de la métrica (que puede ser la traducción de la
finitud o infinitud de la velocidad de propagación de las interacciones) tampoco
determina lo que habitualmente consideramos ``relativista'' ya
que, como hemos visto, la gravitación newtoniana puede presentarse
en dimensión 4 con signatura Minkowski.

La única característica distintiva de los sistemas dinámicos que
comúnmente son llamados relativistas frente a los demás es la
siguiente: \emph{las trayectorias de las partículas de prueba son
parametrizables por el tiempo propio}.

La forma
${\frac{\theta}{\|\theta\|}=\frac{\theta}{\sqrt{|\dot\theta|}}}$
está definida en el abierto de $TM$ en que es  $\dot\theta
(=2T)\ne 0$. En este abierto, la forma es invariante por
homotecias en fibra de $TM$; de hecho, es invariante por todos los
campos tangentes a $TM$ de la forma $\mu\,V$, donde $\mu$ es
cualquier función y $V$ es el generador infinitesimal del grupo de
las homotecias en fibra (en coordenadas locales, $V=\dot
x^i\partial/\partial \dot x^i$ ó $p_i\partial/\partial p_i$). Se
deduce que ${\frac{\theta}{\parallel\theta\parallel}}$ se proyecta
al abierto del espacio de jets $J_1^1M$ (1-jets de curvas en $M$)
imagen del abierto $\dot\theta\ne 0$. La proyección de
$\frac{\theta}{\parallel\theta\parallel}$ es la \emph{forma de
longitud} en $J_1^1M$. Toda curva (no-parametrizada) $\gamma$ en
$M$ define canónicamente una curva en $J_1^1M$; la integral de
$\frac{\theta}{\parallel\theta\parallel}$ sobre esta última es la
longitud de $\gamma$ (habría que precisar: la integral en el
abierto de la curva que queda dentro de la proyección del conjunto
$\dot\theta\ne 0$; al resto de la curva se le adjudica longitud
0). Tiene sentido hablar de longitud de una curva no-parametrizada
porque $\frac{\theta}{\parallel\theta\parallel}$ es proyectable al
espacio de los jets. No ocurre así con la forma de tiempo
$\frac{\theta}{\dot\theta}$: el ``tiempo'' de recorrido de una
curva depende de su parametrización, que la ``sube'' a $TM$.

Salvo detalles técnicos, lo dicho para la longitud se extiende a
todas las medidas de subvariedades de $(M,T_2)$.

 Los físicos llaman \emph{tiempo propio} a la longitud. El tiempo
 propio entre principio y fin de una curva en $M$ tiene sentido
 intrínseco. El tiempo, no. De ahí las ``paradojas de los
 gemelos'', en las que se comparan tiempos propios de dos
 trayectorias distintas (aunque con los mismo extremos) como si
 fueran ``tiempos''.

Dado que la forma que parametriza las trayectorias de  un sistema
mecánico $(M,T_2,\alpha)$ cualquiera es la forma de la clase del
tiempo $\frac{\theta}{\dot\theta}$ (fuera del conjunto
$\dot\theta=0$) y que la forma de longitud o tiempo propio es
$\frac{\theta}{\parallel\theta\parallel}$, el que una trayectoria
del campo sea parametrizable por el tiempo propio significa que en
ella es $\langle
D,\frac{\theta}{\parallel\theta\parallel}\rangle=1$ luego
$\dot\theta={\parallel\theta\parallel}$, es decir, $\dot\theta=1$
ó $\dot\theta=0$. Provisionalmente, impongamos al campo $D$ que
rige la evolución de un sistema relativista la condición de ser
tangente a la hipersuperficie $\dot\theta=1$ de $TM$.

Descompongamos $D$ en suma de campo geodésico y fuerza: $D=D_G+W$;
$D_G$ es siempre relativista porque $D_G\dot\theta=0$; por tanto,
la condición sobre $D$ es $W\dot\theta=0$ en $\dot\theta=1$. Como
$W$ es vertical, es $W\dot\theta=2\langle w,\theta\rangle$, donde
$w=D^\nabla$ es el representante geométrico de $W$. En coordenadas
locales, $\langle w,\theta\rangle=g_{ij}w^i\dot x^j$. Si las $w^i$
son funciones homogéneas de cualquier grado en las $\dot
x^i$,$\langle w,\theta\rangle$ también es homogénea en las $\dot
x$ y, excluyendo el caso en que las $w^i$ sean homogéneas de grado
$-1$ en las $\dot x$, $\langle w,\theta\rangle$ será homogénea de
grado $\ne 0$; luego, si es 0 en la variedad $\dot\theta=1$, es 0
para todos los valores de $\dot\theta$, y será $D\dot\theta=0$
siempre.

 Esta discusión justifica la siguiente
\begin{defi*}
El sistema mecánico $(M,T_2,\alpha)$ es \emph{relativista} si el
campo $D$ que rige su evolución verifica $D\dot\theta=0$,
condición que equivale a que el campo de fuerzas $W=D-D_G$
verifique $W\dot\theta=0$.
\end{defi*}
\begin{thm*}
El sistema mecánico $(M,T_2,\alpha)$ es relativista si y solo si
la forma de trabajo $\alpha$ está en el sistema de contacto
$\Omega$ de $TM$.
\end{thm*}
\begin{proof}[Demostración]
De la ecuación de Newton (\ref{lema0}) se deduce, contrayendo con
$D$: $DT+\dot\alpha=0$, y de aquí
$$D\dot\theta=0\quad\Leftrightarrow\quad
\dot\alpha=0\quad\Leftrightarrow\quad\alpha\in\Omega.$$
\end{proof}

Finalmente resulta que la condición para que un sistema mecánico
sea relativista ¡es independiente de la métrica!

Como el sistema de contacto no contiene más diferencial exacta que
la 0, no hay sistemas conservativos relativistas, salvo el
geodésico. Por eso, no es catalogable como relativista la
gravitación de Newton, mirada en $\mathbb{R}^3$; el que no haya
sistemas conservativos relativistas, salvo el geodésico, no tiene
nada que ver con la in\-exis\-ten\-cia de acción a distancia, o la
finitud de la velocidad de propagación de las interacciones.

Mirando los generadores del sistema de contacto $\Omega$ en
coordenadas locales:
$$\dot x^h dx^k-\dot x^k dx^h=i_{\dot d}(dx^h\wedge dx^k)$$
vemos que, para cada $\alpha\in\Omega$, existe una 2-forma $F_2$
en $TM$ horizontal (combinación lineal de 2-formas en $M$) tal que
$\alpha=i_{\dot d}F_2$: toda fuerza relativista es ``producida''
por un campo tensorial de orden 2, covariante, horizontal,
hemisimétrico, en $TM$. En general, $F_2$ no está unívocamente
determinado por $\alpha$. Cuando el campo de fuerzas $W$ depende
linealmente de las velocidades (es decir, de las $\dot x$),
también es así para $\alpha=-i_wT_2$; en tal caso, existe una única
2-forma $F_2$ en $M$ tal que $i_{\dot d}F_2=\alpha$;
esencialmente, una 2-forma en $M$ es lo mismo que una 1-forma del
sistema de contacto en $TM$ que depende linealmente de las $\dot
x$. Tenemos:
 \begin{thm*}
 Toda 1-forma de fuerza relativista $\alpha$ en $M$ (con cualquier
 métrica) es producida por una 2-forma horizontal $F_2$ de $TM$, mediante la regla
 $\alpha=i_{\dot d}F_2$; cuando se trata de fuerzas que dependen linealmente de las velocidades, la corres\-pon\-den\-cia es canónica y $F_2$ está unívocamente determinada. Una vez dada la métrica $T_2$ en $M$, $\alpha$ determina la fuerza $W=D-D_G$ dada por la ley de Newton (\ref{lema0}), o su representante geométrico $w$, $w=-\textrm{grad}\,\alpha$.
 \end{thm*}

 $W$ es la \emph{fuerza de Lorentz} producida por el tensor $F_2$.

 Independientemente de cual sea la métrica en $M$, no existen
 sistemas conservativos relativistas (salvo el geodésico).

 Compárese este apartado con el tratamiento que hace Barut en \cite{Barut} de las   ``fuerzas de Minkowski''.

 \begin{obs*}[\textbf{Sobre las ``correcciones relativistas''}]
 Dado cualquier sistema me\-cá\-ni\-co $(M,T_2,\alpha)$ se le puede asociar, canónicamente, un sistema relativista $(M,T_2,\overline\alpha)$ imponiéndole la ligadura no-holónoma de tiempo definida por la forma de Liouville $\theta$. El campo $\overline D$ que rige la evolución de $(M,T_2,\overline\alpha)$ está dado por la fórmula (\ref{Campotiempo1}), con $\tau=\theta$. Obsérvese que, en este caso es $\textrm{Grad}\,\theta=V$, el generador infinitesimal del grupo de las homotecias en fibra.

  $\overline D$ es la \emph{corrección relativista} natural (desde el punto de vista general de la Mecánica) del campo $D$. Como ejemplo, puede comprobarse que la fórmula (2.25), Ch. II, de \cite{Barut} se obtiene como resultado de la corrección relativista general que proponemos.
\end{obs*}

\section{Campos electromagnéticos}

Un \emph{campo electromagnético} en la variedad $M$ es una 2-forma
cerrada $F_2$ en $M$. La 1-forma del sistema de contacto
$\alpha=i_{\dot d}F_2$ en $TM$ es la \emph{forma de fuerza de
Lorentz}. Esta forma no depende de la métrica que demos en $M$.
Una vez fijada la métrica, el sistema mecánico $(M,T_s\alpha)$
evoluciona de acuerdo a la ley de Newton (\ref{lema0}); en nuestro
caso
\begin{equation}\label{NewtonLorentz}
i_D\omega_2+dT+i_{\dot d}F_2=0
\end{equation}

Consideramos $F_2$ como una modificación de la forma simpléctica
$\omega_2$ a la nueva forma (también simpléctica) en $TM$:
\begin{equation}\label{Nuevaomega}
\omega_F=\omega_2+F_2.
\end{equation}
que permite escribir la ley de Newton  (\ref{NewtonLorentz}) como:
\begin{equation}\label{NewtonLorentznueva}
i_D\omega_F+dT=0
\end{equation}

El \emph{campo de Lorentz} $D$ es el campo hamiltoniano, de
hamiltoniano $T$, respecto de la estructura simpléctica
$\omega_F$.

La ecuación (\ref{lema01}) de las integrales intermedias da:
\begin{equation*}
i_u(F_2+di_uT_2)+dT(u)=0
\end{equation*}
ó
\begin{equation}\label{intermediaLorentz2}
i_u({\omega_F}\mid_u)+dT(u)=0
\end{equation}
donde ${\omega_F}\mid_u$ es la especialización de la forma
simpléctica $\omega_F$ a la sección $u$, pasada luego por $u^*$ a
$M$.

Si $u$ es una sección de $TM$ lagrangiana para la forma
simpléctica $\omega_F$ (${\omega_F}\mid_u=0$), la ecuación
(\ref{intermediaLorentz2}) da
\begin{equation}\label{intermediaLorentz4}
T(u)=\textrm{cte.}
\end{equation}
Para escribirla como ecuación en derivadas parciales de primer
orden, to\-me\-mos una primitiva local del campo electromagnético
$F_2$; el gradiente de esta primitiva es lo que se denomina un
\emph{potencial vector} de $F_2$: un campo tangente $A$ en $M$ que
verifica:
\begin{equation}\label{potencialvector}
di_AT_2=F_2
\end{equation}
Como
${\omega_F}\mid_u={\omega_2}\mid_u+F_2=di_uT_2+di_AT_2=di_{u+A}T_2$,
la condición para que $u$ sea $\omega_F$-lagrangiana es que
$i_{u+A}T_2$ sea cerrada, luego (localmente) $i_{u+A}T_2=dS$, ó
\begin{equation}\label{ugradiente}
u=\textrm{grad}\,S-A.
\end{equation}
La ecuación (\ref{intermediaLorentz4}) puede escribirse como:
\begin{equation}\label{HJLorentz}
T(\textrm{grad}\,S-A)=\frac 12\, m^2\quad\text{(constante
cualquiera),}
\end{equation}
la \emph{ecuación de Hamilton-Jacobi}, una ecuación en derivadas
parciales de primer orden para la función $S$. En ausencia de campo
electromagnético, (\ref{HJLorentz}) es la ecuación \emph{eikonal}.

Pongamos $\Psi=e^{iS}$. El cálculo hecho antes de la fórmula
(\ref{Schrodinger}), válido para cualquier función $S$ en $M$,
daba
 \begin{equation}\label{DeltaKG}
  \Delta\Psi= [i\Delta S-T_2(\textrm{grad}\,S,\textrm{grad}\,S)]\Psi
  \end{equation}
Sean $S$ una función dada en $M$, $A$ un potencial vector para el
campo electromagnético $F_2$; sea $u=\textrm{grad}\,S-A$.
Sustituyendo en la ecuación (\ref{DeltaKG}), queda
\begin{align*}
\Delta\Psi&=\left[i\,\textrm{div}\,(u+A)-T_2(u+A,u+A)\right]\Psi\\
&=\left[i\,\textrm{div}\,(u+A)-\|u\|^2-\|A\|^2-2\,T_2(A,\textrm{grad}\,S-A)\right]\Psi\\
&=\left[i\,\textrm{div}\,(u+A)-\|u\|^2+\|A\|^2-2\,T_2(A,\textrm{grad}\,S)\right]\Psi\\
&=\left[i\,\textrm{div}\,(u+A)-\|u\|^2+\|A\|^2-2\,A(S)\right]\Psi\\
&=\left[i\,\textrm{div}\,(u+A)-\|u\|^2+\|A\|^2+2i\,A\right]\Psi.\\
\end{align*}
luego
\begin{equation}\label{KG}
\left[\Delta-2i\,A-\|A\|^2+\|u\|^2-i\,\textrm{div}\,(u+A)\right]\Psi=0.
\end{equation}
De esta identidad, con $u=\textrm{grad}\,S-A$, se deduce:
\begin{thm*}
Sean $A$ un potencial vector para el campo electromagnético $F_2$,
$S$ una función en $M$. De las tres condiciones  siguientes, cada
par implica la tercera:
\begin{enumerate}[A)]
\item  $S$ verifica la ecuación de Hamilton-Jacobi
(\ref{HJLorentz}): $$T(\textrm{grad}\, S-A)=\frac 12\,
m^2,\quad\text{constante},$$ lo que es equivalente a decir que
$u=\textrm{grad}\,S-A$ es una integral intermedia de la fuerza de
Lorentz.
\item $S$ es armónica: $\Delta S=0$.
\item $\Psi=e^{iS}$ verifica la \emph{ecuación de Klein-Gordon}:
$$\left(\Delta-2i\,A-\|A\|^2+m^2\right)\Psi=0.$$
\end{enumerate}
\end{thm*}

\noindent\textbf{Observaciones sobre el primer par de ecuaciones
de  Maxwell.} El clásico ``primer par de ecuaciones de Maxwell''
se condensa en los textos actuales en la ecuación
\begin{equation}\label{Maxwell}
\textrm{grad}\,\delta F_2=J\quad\text{ó},\quad \delta
F_2=J^*=i_JT_2.
\end{equation}
Mientras que el ``segundo par de ecuaciones de Maxwell'' $dF_2=0$
no depende de la métrica, el primero sí. Desde el punto de vista
que adoptamos aquí, (\ref{Maxwell}) debe entenderse como
\emph{definición de la corriente eléctrica} $J$ (el ``cuadrivector
de carga-corriente'' clásico) y el problema que se plantea es si $J$
es una integral intermedia del campo de Lorentz determinado por
$F_2$ en $(M,T_2)$. Sin imponer otras condiciones, no; $J$ no es
una corriente de partículas obediente a la fuerza de Lorentz, en
general. Por ejemplo, en $\mathbb{R}^4$-Minkowski puede cambiarse
$F_2$ sumándole un tensor con coeficientes constantes (en
coordenadas vectoriales) sin que cambie $J$, pero cambiando la
fuerza de Lorentz. Sin abordar el problema en general, se tiene:
\begin{thm*}
Si la corriente de Maxwell $J=\textrm{grad}\,\delta F_2$ es
integral intermedia de la fuerza de Lorentz y es
$\omega_F$-lagrangiana , se verifica
\begin{enumerate}[1)]
\item  $-J$ es un potencial vector para $F_2$.
\item $\|J\|$ es constante, $\pm m$, digamos.
\item Si $A$ es un potencial vector que verifica la condición de
\emph{gauge de Lorentz}: $\textrm{div}\,A=0$, y es $J+A=dS$ (la
existencia local de $S$ es consecuencia de 1)), se cumple la
\emph{ecuación de Klein-Gordon} para $\Psi=e^{iS}$:
\begin{equation}\label{KG2}
\left(\Delta-2i\,A-\|A\|^2+m^2\right)\Psi=0.
\end{equation}
\end{enumerate}
\end{thm*}
\begin{proof}[Demostración]
\begin{enumerate}[1)]
\item  Que $J$ sea $\omega_F$-lagrangiano significa que, para cualquier potencial-vector
$A$, es $di_{J+A}T_2=0$, luego $di_{-J}T_2=di_{A}T_2=F_2$.
\item Por ser $J$ integral intermedia del campo de Lorentz y $\omega_F$-lagrangiano, verifica
(\ref{intermediaLorentz4}).
\item La definición de $J^*$ como $\delta F_2$ da automáticamente $\textrm{div}\,J=0$. Luego $\textrm{div}\,(J+A)=0$,
$\Delta S=0$, y se concluye aplicando el teorema anterior.
\end{enumerate}
\end{proof}

\begin{Obs*}
Los tres artículos de Dirac \cite{Dirac} están relacionados con
este apartado. En \cite{Dirac}-I impone al potencial vector la
condición de gauge $\|A\|$=cons\-tan\-te.

Si $F_2$ verifica las condiciones del teorema, es
$$\Delta F_2=d\delta F_2=dJ^*=-F_2,$$
luego
\begin{equation}\label{Fautovalor}
(\Delta+1)F_2=0.
\end{equation}
La existencia de una 2-forma no nula que verifique esta ecuación
impone condiciones a la geometría $(M,T_2)$. Por ejemplo, $T_2$ no
puede ser definida positiva.
\end{Obs*}

\section{Sobre el principio de Hamilton-Noether}

Recordemos que un campo tangente $\delta$ sobre $TM$ es una \emph{transformación infinitesimal de contacto} si $\mathcal{L}_\delta\Omega\subseteq\Omega$ (es decir, para
cada 1-forma
  $\sigma\in\Omega$, la derivada de Lie $\mathcal{L}_\delta\sigma$ está en $\Omega$).

  \begin{thm*} \label{variacion}
  Sea $\delta$ una transformación infinitesimal de contacto en $TM$,
  proyectable a $M$. Las siguientes propiedades son equivalentes:
  \begin{enumerate}
  \item \label{variacion1}
   $\delta$ deja invariante la clase del tiempo: para cada 1-forma
  $\alpha$ en $M$, en el abierto de $TM$ en que $\dot\alpha\ne 0$, tenemos:
  $$\mathcal{L}_\delta\left(\frac{\alpha}{\dot
  \alpha}\right)\equiv 0\,\,\,{\text{\emph{mod}}}\,\, \Omega\ .$$
  \item\label{variacion2}
   $\delta$ conmuta con $\dot d$; es decir, $\dot d\circ\delta =\delta\circ\dot d$
  como derivaciones de $\cC^\infty(M)$ en $\cC^\infty(TM)$.
  \end{enumerate}
  \end{thm*}

  \begin{proof}
  Sea $D$ una ecuación diferencial de segundo orden y  $\alpha$ una
  1-forma en $M$. En el abierto de $TM$ donde $\dot\alpha\ne 0$, tenemos
  $$0=\delta(1)=\delta\left\langle\frac{\alpha}{\dot\alpha},D\right\rangle
           =\left\langle \mathcal{L}_\delta\left(\frac{\alpha}{\dot\alpha}\right),D\right\rangle
            +\left\langle \frac{\alpha}{\dot\alpha},[\delta,D]\right\rangle\ .$$

           El primer término en la última suma es $0$ para
  $D$ arbitrario si y solo si $\delta$ satisface (\ref{variacion1}). El segundo término es $0$ para $\alpha$ arbitrario si y solo si $[\delta,D]$ es vertical, y esto es equivalente a
  (\ref{variacion2}).
  \end{proof}

  \begin{obs*}
   La propiedad (\ref{variacion2})  es la igualdad
  `$d\circ\delta=\delta\circ d$' de los textos clásicos de
  Mecánica; e.g. en Sommerfeld
  \cite{SommerfeldMecanica}, formulas (9) y (9a) de la página 183.
  Nuestra fórmula (\ref{variacion1}) del teorema recuerda a
  ``dejar fijo el tiempo'' en el razonamiento de Sommerfeld en las páginas 183, 184. Sommerfeld atribuye a Euler la fórmula
  $d\delta=\delta d$. Por esta razón, llamamos a  (\ref{variacion2}) del teorema
  \emph{fórmula de conmutación de Euler}.
  \end{obs*}

  \begin{defi*}[\textbf{Variación Infinitesimal}]
  Una transformación infinitsimal de contacto que es proyectable a
  $M$ y satisface la condiciones equivalentes del teorema anterior es llamada una \emph{variación infinitesimal}.
  \end{defi*}

  \begin{thm*}
  Para cada campo $v$ en $M$, existe una única variación infinitesimal $\delta_v$ sobre $TM$ que se proyecta a $M$
  como $v$. 
  \end{thm*}
  \begin{proof} Cuando $\delta_v$ existe, la conmutación de $\dot d$ y $\delta_v$ da
  para cada $f\in\cC^\infty(M)$,
  $\delta_v\dot f=\delta_v\dot df=\dot d\delta_vf=\dot d(vf)\ .$
  Así que $\delta_v$ está determinado por $v$. Recíprocamente, dado en coordenadas
  $v=a^i\partial/\partial x^i$, el campo $a^i\partial/\partial x^i+\dot a^i\partial/\partial\dot x^i$ es una variación infinitesimal (local)
  que se proyecta sobre $v$.
    \end{proof}

    Es fácil comprobar que todas las tranformaciones inifinitesimales de contacto $\delta$ proyectables a un campo dado $v$ en $M$ son de la forma
  \begin{equation}\label{variacionesproyectables}
  \delta=\delta_v+\mu V
  \end{equation}
  donde $V$ es el generador de las homotecias en las fibras de $TM$ y $\mu\in\cC^\infty(TM)$ es  arbitraria.

  El punto de partida para la aplicación de los metodos variacionales en Mecánica es la llamada
   en \cite{Prange},
  ``Zentralgleichung von Lagrange'':

  \begin{thm*} \label{zentral}
    Para cada ecuación diferencial de segundo orden
  $D$ y cada variación infinitesimal $\delta=\delta_v$, se tiene
  \begin{equation}\label{eq:zentral}
  D\langle\theta, \delta\rangle=
    \langle dT-\alpha, \delta\rangle\ ,
  \end{equation}
   donde  $\alpha$ es la forma de trabajo correspondiente a  $D$ por el
   Lema 0.
    \end{thm*}
  \begin{proof}
  La fórmula de Cartan para la derivada de Lie da
  $$
  \mathcal{L}_D\theta =i_D d\theta+d\langle\theta, D\rangle
             =i_D\omega_2+2 dT=-dT-\alpha+2dT=dT-\alpha\ ,$$
  así,
  $$D\langle\theta, \delta\rangle=\langle dT-\alpha, \delta\rangle
            +\langle\theta, [D,\delta]\rangle
            =\langle dT-\alpha, \delta\rangle\ ,$$
  porque $[D,\delta]$  is vertical debido a la fórmula de conmutación de Euler.
  \end{proof}

 \begin{defi*}[\textbf{función Lagrangiana}]
 En un sistema conservativo $(M,T_2,dU)$, la fucnión $L=T-U$ es llamada
  \emph{función lagrangiana} del sistema.
 \end{defi*}

 Aplicando el teorema anterior a un sistema conservativo, obtenemos
 \begin{thm*}[\textbf{Principio de Hamilton-Noether}]\label{hamilton}
 Sea $(M,T_2,dU)$ un sistema me\-cá\-ni\-co conservativo, $D$ la correspondiente ecuación
 diferencial de segundo orden y  $L$ la función lagrangiana. Para cada variación infinitesimal $\delta$, tenemos
 \begin{equation}\label{eq:hamilton}
 D\langle \theta, \delta\rangle=\delta L\ .
 \end{equation}
 \end{thm*}

  La clásica versión integral del principio de Hamilton se obtiene directamente a partir de la ecuación (\ref{eq:hamilton}). Por otro lado,
  cuando $\delta$ es una \emph{simetría} del sistema, en el sentido de que $\delta L=0$, se deduce de esa misma fórmula que $\langle\theta,\delta\rangle$ es constante a lo largo de las trayectorias del sistema mecánico.
  \bigskip

  Como veremos ahora, existe también una versión del principio variacional para sistemas sometidos a una ligadura de
  tiempo.

 Para ello, seleccionamos de entre todas las transformaciones infinitesimales de contacto de $TM$ proyectables a $M$ aquellas que dejan invariante $\dot x^0$: si $v$ es un campo tangente en $M$, se deduce de  (\ref{variacionesproyectables}) que existe una única transformación $\overline\delta$ proyectable en $v$ y tal que $\overline\delta\dot x^0=0$, a saber,
 \begin{equation}\label{tictiempo}
 \overline \delta=\delta_v+\mu\, V,\quad\text{con}\quad \mu=-\frac{\dot v^0}{\dot x^0}\quad\text{siendo}\quad v^0:=v(x^0)
 \end{equation}

 Calculemos $\overline D\langle \overline\theta,\overline\delta\rangle=\langle \mathcal{L}_{\overline D}\overline\theta,\overline\delta\rangle +
     \langle\overline\theta,[\overline D,\overline\delta]\rangle$, donde $\overline\theta$ está definida por (\ref{Liouvilletiempo}). La relación (\ref{Newtontiempo3}), junto con la fórmula de Cartan para la derivada de Lie y la igualdad $i_{\overline D}\overline\theta=L$, nos dá
     $$\mathcal{L}_{\overline D}\theta=dL-\frac H{\dot x^0}d\dot x^0,$$
     así que $$\langle \mathcal{L}_{\overline D}\overline\theta,\overline\delta\rangle=\overline\delta L$$ porque $\overline\delta\dot x^0=0$.
 Por otra parte, es fácil ver utilizando (\ref{tictiempo}) que
$$[\overline D,\overline\delta]\equiv -\mu \, \dot d,$$
módulo componentes verticales (que podemos obviar porque se anularán al aplicar $\overline\theta$).
Contrayendo,
$$\langle  \overline\theta, [\overline D,\overline\delta]\rangle=
    \left\langle \theta-H\frac{dx^0}{\dot x^0},\frac{\dot v^0}{\dot x^0}\dot d\right\rangle=(2T-H)\frac{\dot v^0}{\dot x^0}=L\frac{\dot v^0}{\dot x^0}
;$$
sumando queda
\begin{equation}\label{bla}
\overline D\langle\overline\theta,\overline\delta\rangle=\overline\delta L+L\frac{\dot v^0}{\dot x^0}.
\end{equation}
Podemos interpretar adecuadamente el segundo miembro  observando que
$$\mathcal{L}_{\overline\delta}(L\,dx^0)=\overline\delta(L)\,dx^0+L\,d(\overline\delta x^0)=\overline\delta(L)\,dx^0+L\,dv^0,$$
y que, además,
$$dv^0\equiv\frac{\dot v^0}{\dot x^0}\,dx^0\quad ({\textrm{mod}}\,\,\Omega);$$
luego,
\begin{equation*}
\mathcal{L}_{\overline\delta}(L\,dx^0)\equiv \left(\overline\delta L+L\frac{\dot v^0}{\dot x^0}\right)\,dx^0\quad ({\textrm{mod}}\,\,\Omega).
\end{equation*}
Sustituyendo en la ecuación (\ref{bla}), después de multiplicarla por $dx^0$, resulta el

 \begin{thm*}[\textbf{Principio de Hamilton-Noether para una ligadura de tiempo}]
 Sea $(M,T_2,dU)$ un sistema me\-cá\-ni\-co conservativo, $D$ la correspondiente ecuación
 diferencial de segundo orden y  $L$ la función lagrangiana. Sean, además, $\overline\theta=\theta-(H/\dot x^0)dx^0,$ $\overline D$ el campo $D$ modificado por la
  ligadura de tiempo para $\tau=dx^0$ y $\overline\delta$ una transformación infinitesimal  proyectable a $M$ y tangente a dicha ligadura. Entonces se tiene
  \begin{equation}\label{Hamiltontiempo}
\overline D\langle\overline\theta,\overline\delta\rangle\,dx^0\equiv \mathcal{L}_{\overline\delta}(L\,dx^0)\quad ({\textrm{mod}}\,\,\Omega);
\end{equation}
relación que se mantiene por restricción a la subvariedad de ligadura $\dot x^0=1$ (o, más en general, a $\dot x^0=\textrm{cte.}\ne 0$).
 \end{thm*}
\smallskip

Consideraciones análogas a las hechas para (\ref{eq:hamilton}) son aplicables igualmente a (\ref{Hamiltontiempo}), dado que el sistema de contacto $\Omega$ se anula sobre las trayectorias.

\begin{obs*}
La subvariedad $\dot x^0=1$ es una sección local de la proyección de $TM$ sobre el espacio de 1-jets de curvas $J_1^1(M)$ y por tanto se identifica con éste. Es fácil comprobar que si $\overline \delta$ se proyecta sobre $v$ en $M$, entonces, bajo la mencionada identificación, la restricción de $\overline\delta$ no es otra cosa que la  prolongación del campo $v$ a $J_1^1M$.
\end{obs*}

\section{La ecuación de Schrödinger}

La ecuación de Schrödinger que aparece en el teorema de la sección 1 se deduce de la de Hamilton-Jacobi y de la hipótesis, muy restrictiva, de la conservación de la forma de volumen por el campo del movimiento; consecuencia de lo estrcito de las hipótesis es que las soluciones son de amplitud constante.

En esta sección volvemos sobre el tema de la relación de Hamilton-Jacobi con Schrödinger, pero partiendo de las condiciones más generales dentro de lo que parece físicamente razonable. Por eso, creemos que la ecuación de Schrödinger ``clásica'' a que llegamos está a la mínima distancia posible de la ecuación cuántica, aunque solo hemos empezado a explorar casos en que coinciden o no.

\subsection*{La acción como fase}
En un sistema mecánico conservativo $(M,T_2, dU)$ se llama \emph{acción} a cualquier solución de la ecuación de Hamilton-Jacobi $H(\textrm{grad}\, S)=E$. Tales soluciones se construyen partiendo de una subvariedad (localmente cerrada) $X$ de $M$, considerando su fibarado conormal $X^*\subset T^*M$, cortándolo por la hipersuperficie $H=E$ y propagando la subvariedad $(n-1)$-dimensional resultante, $X_E^*$, por el campo hamiltoniano $D$ (pasado a $T^*M$, o pasando $X_E^*$ a $TM$). Por su propia definición, la forma de Liouville $\theta$ se especializa a $X^*$ como  0; como $D$ es el radical de la especialización de $\omega_2=d\theta$ a $H=E$, la propagación de $X_E^*$ por $D$ dá una variedad lagrangiana $\Sigma$ de $T^*M$. En esta variedad $\Sigma$, $\theta$ es localmente exacta; existe una función $S$ en (un abierto de) $M$ tal que $\theta|_\Sigma=dS$; esta función puede normalizarse por la condición de ser $S=0$ en $X_E^*$.

En los textos suele tomarse como variedad inicial $X$ un punto $a$ de $M$; con ello, $S$ es (en un entorno de $a$) la integral de $\theta$ a lo largo de las trayectorias que parten de $a$, con energía dada $E$.

En cada subvariedad $S=\textrm{cte}$ de $\Sigma$, la $\theta$ se especializa como 0; genéricamente (no es preciso detallar), $S=\textrm{cte}$ es una variedad de la forma $Y^*$, con $Y$ una hipersuperficie en $M$. La forma de longitud $\theta/\|\theta\|$ se especializa como 0 en $Y^*$: \emph{la distancia entre dos puntos $\alpha$, $\beta$ de $Y^*$, medida a lo largo de curvas en $Y^*$ es 0, aunque $Y$ no sea un punto}. Por eso es obligado considerar $Y_E^*$ como un frente de onda y $S$ como la fase de cualquier cosa que consideremos como \emph{ondas de materia} asociadas al sistema mecánico y a la solución $\Sigma$. Dentro de cada frente de onda y medido en tiempo propio, la duración del  paso de un estado a otro es cero.

\subsection*{La amplitud de una onda}
La fase $S$ determina la energía $E=H(dS)$ (la ``frecuencia'') y la velocidad de propagación $u=\textrm{grad}\, S$ (salvo un factor). La intensidad, ``energía'' de la onda, amplitud..., son otra cosa que la $E$. Como ocurre con las ondas sonoras, la intensidad no hace cambiar la frecuencia ni la velocidad de propagación. De cualquier modo que tratemos de interpretar la amplitud de una onda (densidad de energía, probabilidad de presencia, etc.) debe estar relacionada con algo que sea conservado por el flujo $u$. Cuando imponíamos la condición $\textrm{div}\,u=0$ (conservación del volumen), obteníamos como condición la ecuación de Schrödinger para funciones de onda de amplitud constante. Impongamos la condición de que $u$ conserve una $n$-forma $\rho\, \omega_n$ ($\omega_n$ es la forma de volumen); la condición es $u(\rho)+\rho\,\textrm{div}\,u=0$, que es
$$T_2(\textrm{grad}\, S,\textrm{grad}\,\rho)+\rho\,\Delta S=0,\quad\text{ó}$$

$$u(\textrm{log}\,\rho)+\Delta S=0\quad\text{(\emph{condición de conservación}).}$$

Esta ecuación muestra que las posibles amplitudes en la variedad lagrangina $u=\textrm{grad}\,S$ están determinadas salvo un factor, que es una integral primera del flujo $u$.

\subsection*{Ondas en $u=\textrm{grad}\,S$}
Elegida la función $\rho$ de modo que el flujo $u$ conserve la densidad $\rho\,\omega_n$, estudiemos la ecuación que satisface la función de onda
$$\Phi=f(\rho)\,e^{iS}$$
donde $f$ es una función, de momento indeterminada. Haciendo los cálculos ha\-bi\-tua\-les, queda la identidad:
\begin{multline*}
 \phantom{mm}\Delta\Phi=\left\{\frac{f'(\rho)}{f(\rho)}\,\Delta\rho+\frac{f''(\rho)}{f(\rho)}\|\textrm{grad}\,\rho\|^2-\|\textrm{grad}\,S\|^2+\right.\\
     +\left.i\left[\Delta S+2\,\frac{f'(\rho)}{f(\rho)}\,T_2(\textrm{grad}\,S,\textrm{grad}\,\rho)\right]\right\}\,\Phi\phantom{mm}.
\end{multline*}
Usando la condición de conservación, la parte imaginaria de $\{\phantom{i}\}$ es
$$\Delta S\left(1-\frac{2\rho f'(\rho)}{f(\rho)}\right).$$
Precindiendo de la hipótesis de que $S$ sea armónica, la anulación de tal parte imaginaria exige que, salvo un factor, sea $f(\rho)=\sqrt{\rho}$ ó $\rho=|\Phi|^2$, que es consistente con las interpretaciones habituales de $\rho$.

Tomando así la $f$, queda
\begin{equation*}
\left[\Delta+\|\textrm{grad}\,S\|^2\right]\Phi=\left[\frac{\Delta \rho}{2\rho}-\frac 1{4\rho^2}\|\textrm{grad}\,\rho\|^2\right]\Phi
  =\frac{\Delta\sqrt{\rho}}{\sqrt{\rho}}\,\Phi,
\end{equation*}
luego
$$\left(-\frac 12\,\Delta+U\right)\,\Phi=\left(H-\frac{\Delta\sqrt{\rho}}{2\sqrt{\rho}}\right)\,\Phi,$$
de donde se deduce:
\begin{thm*}
Sean $(M,T_2,dU)$ un sistema conservativo, $S$ una función en $M$, $u=\textrm{grad}\,S$, $\rho$ una función tal que $\mathcal{L}_u(\rho\,\omega_n)=0$. De las tres condiciones siguientes, cada par implica la tercera:
\begin{enumerate}[A)]
\item  $S$ verifica la ecuación de Hamilton-Jacobi
 $$H(dS)=E.$$
\item $\sqrt{\rho}$ es armónica.
\item $\Phi=\sqrt{\rho}\,e^{iS}$ verifica la ecuación de Schrödinger
$$\left(-\frac 12\Delta+U\right)\,\Phi=E\,\Phi.$$
\end{enumerate}
\end{thm*}

En relación con este apartado, comparar con los puntos 2.6, 2.7 en Holland \cite{Holland}.

La versión más general de este teorema se obtiene admitiendo como amplitud una función compleja $a$:
\begin{thm*}
Sean $(M,T_2,dU)$ un sistema conservativo. Sean $S$ una función real, $a$ una función compleja en $M$. De las tres condiciones siguientes, cada par implica la tercera:
\begin{enumerate}[A)]
\item  $S$ verifica la ecuación de Hamilton-Jacobi
 $$H(dS)=E.$$
\item $\textrm{div}\,(a^2\,\textrm{grad}\,S)=ia\,\Delta a$.
\item $\Phi=a\,e^{iS}$ verifica la ecuación de Schrödinger
$$\left(-\frac 12\Delta+U\right)\,\Phi=E\,\Phi.$$
\end{enumerate}
\end{thm*}
La demostración es la ya acostumbrada.

Dejamos para otra ocasión el tratamiento de las ecuaciones de Schrödinger con tiempo y de Klein-Gordon.
\bigskip


\end{document}